%% file: main.tex
\documentclass[%
 reprint,
 amsmath,amssymb,
 aps,
]{revtex4-2}
\usepackage{comment}
\usepackage{graphicx}
\usepackage{dcolumn}
\usepackage{bm}

\usepackage{footnote}
\usepackage{url}
\usepackage{hyperref}
\usepackage{multirow}
\begin{document}

\preprint{APS/123-QED}

\title{Classical Autoencoder Distillation of Quantum Adversarial Manipulations}

\author{Amena Khatun\textsuperscript{1}}
\email{amena.khatun@data61.csiro.au}
\author{Muhammad Usman\textsuperscript{1,2}}

\affiliation{%
\textsuperscript{1}Quantum Systems, Data61, CSIRO, Australia\\
\textsuperscript{2}School of Physics, The University of Melbourne, Victoria, Australia}%

\begin{abstract}
Quantum neural networks have been proven robust against classical adversarial attacks, but their vulnerability against quantum adversarial attacks is still a challenging problem. Here we report a new technique for the distillation of quantum manipulated image datasets by using classical autoencoders. Our technique recovers quantum classifier accuracies when tested under standard machine learning benchmarks utilising MNIST and FMNIST image datasets, and PGD and FGSM adversarial attack settings. Our work highlights a promising pathway to achieve fully robust quantum machine learning in both classical and quantum adversarial scenarios. 
\end{abstract}

\maketitle

\label{intro}
Machine learning (ML) is rapidly becoming ubiquitous and its integration in real-world applications is already addressing highly complex challenges such as early disease detection using medical imaging \cite{rana2023machine, varoquaux2022machine}, anomaly detection \cite{ahmed2016survey}, natural language processing for chatbots \cite{young2018recent},  image recognition for autonomous vehicles \cite{fujiyoshi2019deep}, and climate modeling for environmental conservation \cite{reichstein2019deep}. Despite efficiency and high throughput of ML systems, there is a serious concern about their vulnerability to data spoofing or manipulations \cite{szegedy2013intriguing,huang2011adversarial,goodfellow2014explaining,kurakin2018adversarial,ilyas2018black,tjeng2017evaluating}, in particular for security-sensitive applications such as self-driving vehicles \cite{ibrahum2025deep} and military systems \cite{lee2024evasion, chen2022risk}. Meanwhile, the developments in the field of quantum ML have instigated a new line of research in which the vulnerability of quantum ML models is being tested and benchmarked in the recent literature \cite{lu2020quantum, ren2022experimental, gong2022universal, wendlinger2024comparative, anil2024generating, winderl2024quantum, guan2021robustness, liu2020vulnerability, west2023benchmarking}. It has been shown that while quantum ML models are robust against classical adversarial attacks, they have been found vulnerable against quantum adversarial attacks - data manipulations generated via quantum models \cite{liu2020vulnerability, west2023benchmarking}. Therefore, techniques that can restore trust and reliability in quantum ML models will have important implications for future practical applications of these models.    
\begin{figure*}
\begin{center}
\includegraphics[width=1.0\linewidth]{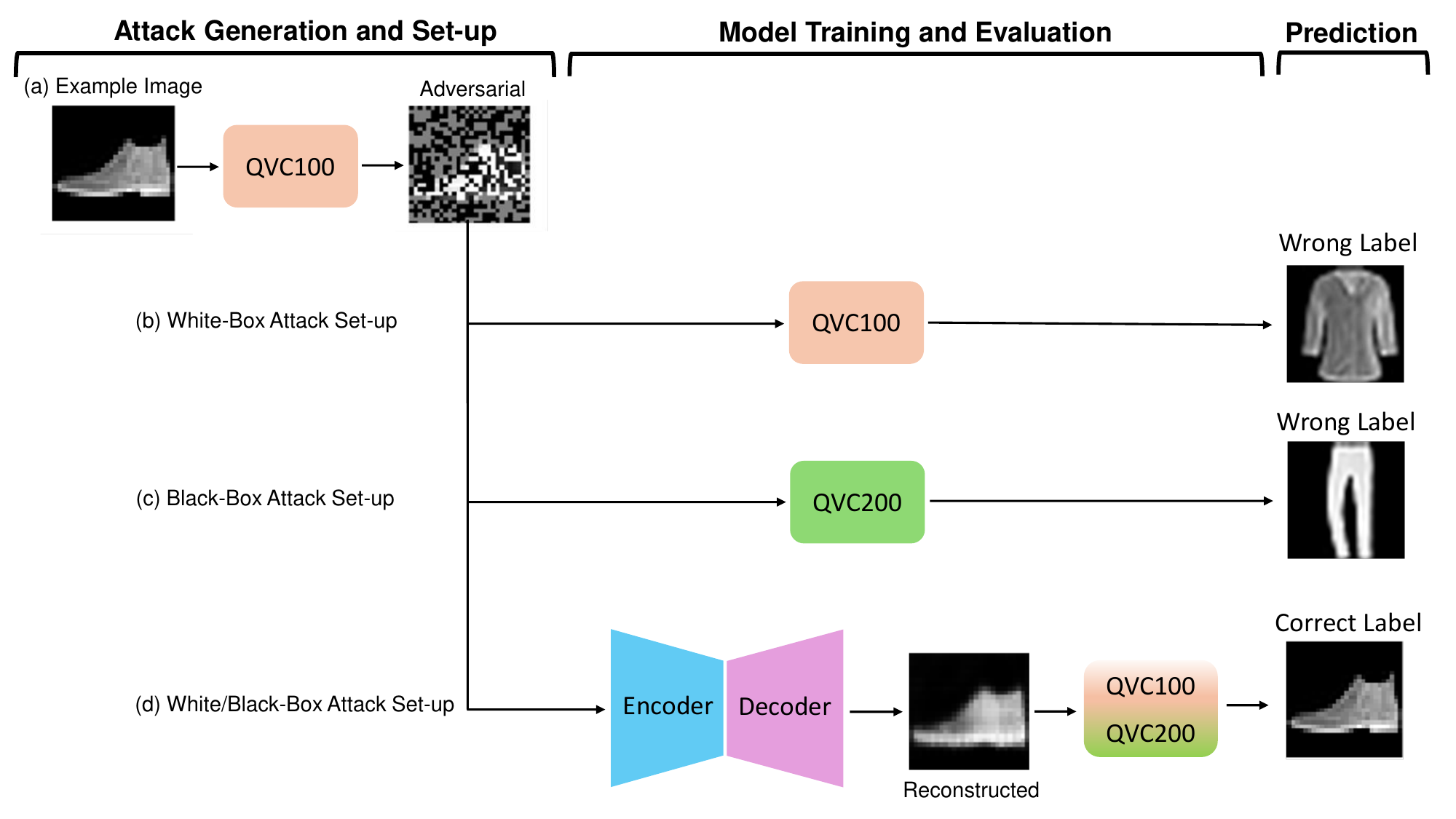}
\end{center}
   \caption{Overview of the proposed QVC-CED approach. In this figure, (a) represents the first step in our workflow where clean images are input to a quantum variational circuit to generate adversarial samples using gradient-based perturbations (PGD or FGSM). (b) and (c) depicts that the perturbed images are passed to quantum classifiers under white-box and black-box attack scenarios, respectively, to test the vulnerability of the classifiers against quantum adversarial attacks. (d) represents that the perturbed images are first fed into a classical encoder-decoder model to purify and reconstruct the original images from the noisy images. After the reconstruction, classification is performed again in both white-box and black-box settings to test the performance of the autoencoder in counteracting adversarial effects.}
\label{fig:overview_main}
\end{figure*}

This work proposes and implements a new quantum-classical framework which integrates a quantum variational classifier with a classical encoder-decoder (QVC-CED) model paving the way for an end-to-end architecture with robustness against both classical and quantum attacks. In classical ML literature, autoencoders have been widely used for denoising of images for classical neural networks \cite{bajaj2020autoencoders, vincent2008extracting}, however our work is the first to integrate it with a quantum neural network and demonstrate its effectiveness for the distillation of quantum adversarial noise. In the context of quantum ML and adversarial perturbations, previous studies have explored various methods to protect quantum ML models against adversarial attacks which include quantum adversarial training  \cite{khatun2024quantum, lu2020quantum, ren2022experimental, montalbano2024quantum},
quantum noise-based defense \cite{du2021quantum, winderl2024quantum} and randomized
encoding \cite{gong2024enhancing}. Other studies \cite{west2023benchmarking, west2023towards, west2024drastic, wu2023radio, dowling2024adversarial}  have investigated intrinsic quantum properties serve as natural defenses against adversaries. A detailed description and discussion of the recent relevant literature is provided in the supplementary section \ref{background}. Previous work on defensive approaches offer limited scope due to their implementation on either down-scaled MNIST data \cite{lu2020quantum, kehoe2021defence}, or simple binary classification problem \cite{ren2022experimental}. Our work is the first to demonstrate that robust implementation is possible for full-scale MNIST and FMNIST datasets under complex projected gradient descent (PGD) \cite{madry2017towards} and fast gradient sign method (FGSM) \cite{goodfellow2014explaining} attacks which are standard benchmarks for ML literature.

Figure \ref{fig:overview_main}  schematically illustrates our quantum-classical approach in which quantum adversarial attacks are generated via a quantum variational circuit which are then passed to a pre-trained quantum classifier. The adversarial images are able to trick quantum classifier to mislabel datasets. A classical encoder/decoder architecture purifies quantum adversarial images and restore the classification accuracy of  the quantum classifier. In supplementary document \ref{sup-QVC} and \ref{sup-enc}, we provide detailed circuit diagrams for quantum classifier and the architecture of classical encoder/decoder circuit.

\label{lit_review}

The validation of our technique is performed by implementing the proposed QVC-CED framework using the Pennylane and PyTorch software frameworks. The QVC model was initialised with 10 qubits and a depth of 100/200 parameterised layers, offering sufficient expressive power for complex feature extraction. The model was trained using the Adam optimizer with a learning rate of 0.005 and a batch size of 256 for 20 epochs. Cross entropy loss was employed as the loss function to train the QVC classifier. To simulate quantum adversarial attacks, we generated gradient-based perturbations using the  FGSM and PGD. The details of FGM and PGD implementations are in supplementary section \ref{Sup-attack-methods}. The classical autoencoder, comprising a CNN-based encoder and a decoder with transposed convolution layers, was trained using adversarially perturbed images as input and their clean counterparts as targets. The autoencoder was trained with mean squared error loss and the Adam optimizer with a learning rate of 0.001 for 20 epochs on the same datasets. All experimental simulations were conducted using a high-performance computing system equipped with a single NVIDIA GPU, enabling efficient simulation of hybrid quantum-classical workflows.

In this work, we considered MNIST and FMNIST datasets without any pre-processing or downscaling, therefore our benchmarks are at par with standard approaches adapted by classical literature \cite{goodfellow2014explaining, kurakin2018adversarial, west2023benchmarking}. MNIST consists of 70,000 grayscale images of handwritten digits ranging from 0 to 9, each represented as a $28\times28$ pixel array. The dataset is split into 60,000 training images and 10,000 test images, with each image labeled according to its corresponding digit. Fashion-MNIST (FMNIST) is a more challenging dataset, providing a robust benchmark for ML and computer vision algorithms. It consists of 70,000 grayscale images of fashion items across 10 categories, such as T-shirts, trousers, dresses, and shoes, with each image labeled according to its respective class. Similar to MNIST, the images are $28\times28$ pixels in size. The dataset is divided into 60,000 training images and 10,000 test images. For both datasets, we evaluate the performance across all 10 classes for the classification task.

\begin{figure*}
\begin{center}
\includegraphics[width=1.0\linewidth]{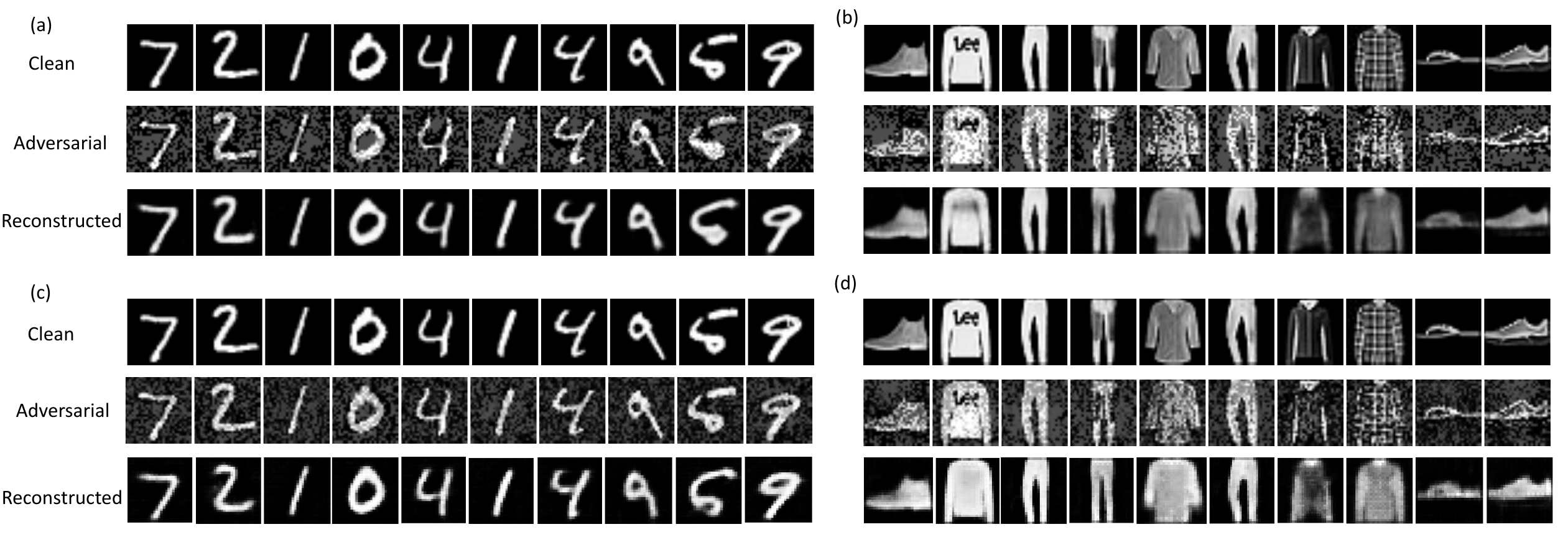}
\end{center}
   \caption{Examples of the original images, adversarially perturbed images and the corresponding reconstructed images, presented in the first, second and third rows, respectively. For all the cases, the results are demonstrated using an attack strength of 0.3, generating strong adversarial perturbations to evaluate the  performance of autoencoder under challenging conditions. (a) and (b) illustrate the adversarial images using FGSM attack on the QVC100 model for both MNIST and FMNIST datasets. (c) and (d) depict the adversarial images generated using PGD attack on the same model. The reconstructed images in the third row, generated by the autoencoder, effectively mitigate the adversarial noise, showcasing its robustness in reconstructing original representations.}
\label{fig:result1}
\end{figure*}

To test the adversarial vulnerability of the QVC model against quantum attacks, we conduct both white-box and black-box attacks. In white-box scenario, the same QVC model is used for both adversarial example generation and evaluation, assuming that the attacker has access to the network.  Specifically, adversarial perturbations are crafted using the QVC100 model and tested on the same model. This white-box testing framework allows to understand the model's resilience against attacks that are specifically designed for it. We generate adversarial examples using PGD and FGSM attacks. Figure \ref{fig:result1} illustrates the examples of the original, adversarially perturbed, and reconstructed images for both datasets. The first, second and third rows show the original images, adversarial examples, and the reconstructed images obtained from the classical autoencoder. In this Figure, we highlight the results for an attack strength of 0.3, which generates strong adversarial perturbations. Figure \ref{fig:result1}(a) and (b) demonstrate the performance under FGSM attack for MNIST and FMNIST datasets, respectively while (c) and (d) represent the results with PGD attack for both datasets. For both PGD and FGSM attacks, the adversarial examples are generated by our QVC100 model that are often imperceptible to the human eye but significantly degrade the classifier’s accuracy. These perturbations are strategically crafted to mislead the classifier while maintaining the visual appearance of the input. At higher attack strengths, these distortions become more prominent, leading to visible artifacts in the adversarial examples.

The classical autoencoder is designed to denoise adversarially perturbed images and restore them to their original appearance. The details about autoencoder are presented in the supplementary section \ref{sup-enc}. Adversarial perturbations shift inputs away from the distribution of clean images, introducing subtle yet targeted changes that can mislead classifiers. Autoencoder focus on reconstructing only the most salient and representative features of the images, tend to ignore these off-manifold adversarial components during reconstruction. This effect aligns with the manifold hypothesis, where the original clean images lie on a low-dimensional surface embedded within the high-dimensional input space \cite{alain2014regularized}. Adversarial examples move inputs off this surface. By projecting inputs back onto the data manifold, autoencoder effectively remove adversarial noise and restore semantic fidelity \cite{vincent2008extracting}. As illustrated in Figure \ref{fig:result1}, the autoencoder significantly reduces the impact of adversarial noise, allowing the reconstructed images to closely resemble their clean counterparts, even under strong attack conditions ($\epsilon=0.3$). For instance, from the reported results under FGSM and PGD attacks, the quantum classifier is prone to misclassification. However, after passing through the autoencoder, the reconstructed images retain essential features, mitigating adversarial distortions and preserving classification integrity.

\begin{figure*}
\begin{center}
\includegraphics[width=0.8\linewidth]{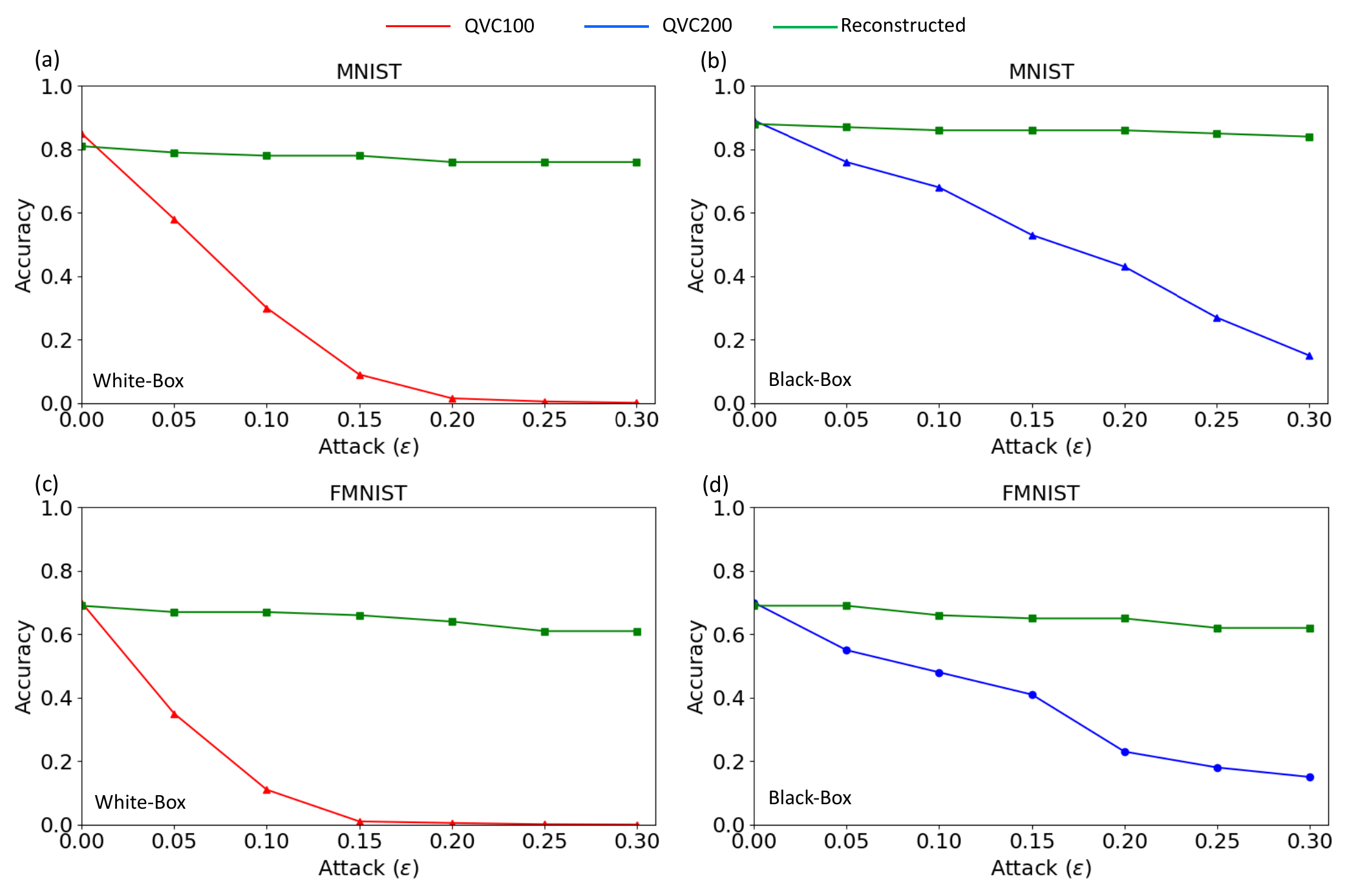}
\end{center}
   \caption{Plot of classification accuracy of quantum classifier under varying attack strengths, ranging from 0 to 0.3. For all the cases, the attacks are generated by QVC100 model. Panels (a), and (c) show the performance of quantum classifier under PGD attacks for the MNIST and FMNIST datasets in white-box setting, respectively.  (b) and (d) represents the impact of PGD attacks on the quantum classifier for the same datasets in black-box scenario. We also report the reconstruction accuracy when there is no attack ($\epsilon=0$). Across all cases, the classification accuracy decreases as the attack strength increases, highlighting the susceptibility of the models to quantum adversarial attacks. However, the results demonstrate a significant recovery in classification accuracy when the adversarial examples are passed through the classical autoencoder.}
\label{fig:result3}
\end{figure*}

To test the resilience of the quantum classifier, we perform the classification task on clean (when the classifier is not under attack), under adversarial attacks, and after defense across all 10 classes of MNIST and FMNIST datasets. Figure \ref{fig:result3} shows the classification accuracies of QVC model under PGD attack in white-box and black-box scenarios. PGD is a more iterative and powerful attack method than FGSM. The attack strength ($\epsilon$) is varied from 0 to 0.3 to test the severity of adversarial impact. In the white-box scenario, where the attacker has full access to the QVC100 model, accuracy plummets as 
$\epsilon$ increases. For MNIST (Figure \ref{fig:result3}(a)), the QVC model achieves an initial accuracy of 85\%. For FMNIST (Figure \ref{fig:result3}(c)), which is more complex as FMNIST images exhibit higher intra-class variability and inter-class similarity,  making them harder to distinguish, the initial accuracy is slightly lower, which is $70\%$. We also report the reconstruction accuracy when there is no attack, i.e., the value of $\epsilon$ is 0. Under PGD attacks, as $\epsilon$ increases, the accuracy of the QVC model drops sharply. For MNIST, accuracy falls from 85\% to nearly 0 at $\epsilon=0.3$. A similar trend is observed for FMNIST, where the accuracy decreases more rapidly, indicating higher vulnerability due to the increased complexity of fashion items compared to handwritten digits. The black-box scenario (Figure \ref{fig:result3}(b), (d)) tests the resilience of quantum classifier when adversarial examples generated by QVC100 model are evaluated on QVC200 model, simulating a realistic attack with limited knowledge of the model. For MNIST, the accuracy decreases from 89\% to 15\% at attack strength of 0.3, a slightly lower drop than in the white-box case, suggesting that black-box attacks are less tailored but still effective. FMNIST dataset also shows a comparable trend in black-box setting. These results highlight that adversarial perturbations exhibit strong transferability across quantum models, even without having any prior knowledge of the target model.
We also report the classification accuracy for both datasets under FGSM attack in the supplementary document (see Figure \ref{fig:result2}). Under FGSM attacks, the model’s robustness deteriorates like PGD attacks, highlighting its susceptibility to adversarial perturbations. The reconstruction accuracy, measured as the classifier’s performance on autoencoder-reconstructed images, remains significantly higher than on adversarial examples. At strength $\epsilon=0.3$, where adversarial noise is at its peak, the classifier’s accuracy on adversarial images drops to 0\% for both MNIST and FMNIST datasets in white-box settings. However, after passing through the autoencoder, the classification accuracy is substantially recovered to 80\% for MNIST and 65\% for FMNIST.

The proposed QVC-CED approach of mitigating adversarial noise plays a significant role in enhancing the performance and reliability of quantum classifier. While quantum attacks expose the vulnerabilities of quantum classifiers, the classical autoencoder effectively mitigates these threats even under strong attack scenarios, reinforcing model reliability. We note that our work has not considered the impact of hardware noise which is present in the current generation of quantum processors and therefore is aimed for the fault-tolerant quantum computing regime. The adversarial attacks investigated in our work are based on standard classical PGD and FGSM strategies implemented through quantum variational circuits. In future, it would be interesting to develop attack strategies which are based on quantum principles such as reported in Ref. \cite{akter2024quantum} which is still an open research question. Finally, another interesting line of research will be to integrate our encoder/decoder framework with quantum machine learning architectures beyond variational approaches such as quantum convolutional neural networks \cite{cong2019quantum} and non-unitary approaches \cite{heredge2024non}. Nevertheless, the successful integration of classical denoising techniques with quantum classifiers reported in this work demonstrates the potential for hybrid quantum-classical architecture that leverage the strengths of both paradigms. Such methods open up possibilities for developing more robust QML models for real-world deployment where adversarial security is a critical concern.
\vspace{4mm}
\section*{Data Availability Statement}
All datasets generated in this work are available in Figures. Further information can be provided upon reasonable request to the corresponding author.

\section*{Acknowledgments}
A.K. acknowledges the use of CSIRO HPC (High-Performance Computing) for conducting all the experimental simulations. A.K. and M.U. also acknowledge CSIRO’s Quantum Technologies Future Science Platform for providing the opportunity to work on quantum machine learning.

\section*{Author Contribution Statement}
M.U. conceived the idea and supervised the project. A.K. developed and implemented the QVC-CED framework and generated all datasets. M.U. and A.K. analysed the data and wrote the manuscript.
\bibliographystyle{unsrt} 
\bibliography{apssamp}
\clearpage
\input{supplementary/sup}

\end{document}

%% file: supplementary/sup.tex

\makeatletter
\renewcommand \thesection{S\@arabic\c@section}
\renewcommand\thetable{S\@arabic\c@table}
\renewcommand \thefigure{S\@arabic\c@figure}
\makeatother

\begin{center}
   {\Large{\textbf{Supplementary Document}}}
\end{center}

\section{Related Work}
\label{background}
The susceptibility of QML models to adversarial attacks have been studied in Ref. \cite{lu2020quantum, ren2022experimental, gong2022universal, wendlinger2024comparative, anil2024generating, winderl2024quantum, guan2021robustness, liu2020vulnerability}. These studies reveal that even subtle modifications to the original input data cause high-performing quantum classifiers to falter under manipulation. These investigations encompasses a wide range of scenarios, including the classification of handwritten digits in the MNIST dataset  \cite{lu2020quantum, ren2022experimental, guan2021robustness}, identification of the phases \cite{lu2020quantum, guan2021robustness}, and classification of quantum data \cite{lu2020quantum, ren2022experimental}. Few existing works have focused on addressing these vulnerabilities through defense strategies. Quantum adversarial training has been demonstrated as an effective approach to improve the robustness against adversarial attacks, as explored in Refs. \cite{khatun2024quantum, lu2020quantum, ren2022experimental, wendlinger2024comparative, montalbano2024quantum}. This method involves retraining quantum classifiers with adversarial examples, which enhances their ability to resist such manipulations. The impact of quantum noise as a natural defense mechanism has been studied extensively in Refs. \cite{du2021quantum, huang2023certified, huang2023enhancing, winderl2024quantum}. These approaches demonstrated that noise inherent to quantum systems can mitigate adversarial perturbations, effectively increasing the robustness of QML models. Randomized and approximate encoding schemes have been proposed in Ref. \cite{gong2024enhancing, west2024drastic} to obscure the decision boundaries of quantum classifiers, thereby reducing their susceptibility to attacks. On the other hand, the theoretical foundations for generating universal adversarial perturbations for QML models have been laid out in Refs. \cite{gong2022universal, anil2024generating}. These studies highlight how adversarial examples can be constructed to mislead quantum classifiers across a wide range of inputs. However, another study, Ref. \cite{dowling2024adversarial}, have shown that quantum circuits exhibiting sufficiently quantum chaotic behavior can inherently resist universal adversarial perturbations. Ref. \cite{west2023benchmarking} systematically demonstrated that quantum classifiers, particularly QVCs, exhibit greater resilience to adversarial attacks compared to their classical counterparts as QVCs learn a distinct yet highly meaningful set of features compared to classical classifiers. Despite these advancements, there remains a significant gap in translating theoretical insights into practical implementations that can reliably defend QML models against adversarial attacks. While the outlined approaches offer promising directions, the development of comprehensive, scalable, and efficient defense mechanisms for real-world QML applications is still an open challenge.

\section{Adversarial Attack Methods}
\label{Sup-attack-methods}
Adversarial attacks are designed to manipulate the predictions of ML models, particularly classifiers. Let us consider a classifier \( f: X \rightarrow Y \), which maps input data \( X \subseteq \mathbb{R}^d \) to a set of discrete output labels \( Y \). The goal of an adversarial attack is to generate a perturbed example \( x^\wedge \) that is almost indistinguishable from a legitimate input \( x \) measured by a small perturbation \( \|x^\wedge - x\| \) while causing the classifier to misclassify the adversarial example. Specifically, the attack ensures that the classifier's prediction changes, i.e., \( f(x^\wedge) \neq f(x) \). When the attacker goal is to induce misclassification without specifying a particular target label for \( f(x^\wedge) \), it is called a \textit{non-targeted attack}. In contrast, a \textit{targeted attack} aims to mislead the classifier into predicting a specific target label \( y \) for \( x^\wedge \), such that \( f(x^\wedge) = y \). Regardless of the type of attack, the adversarial perturbation is designed to remain minimal, ensuring \( x^\wedge \) stays perceptually similar to \( x \). The similarity or distance between \( x \) and \( x^\wedge \) is typically measured using the \( L_p \) norm. Different \( p \)-norms capture distinct ways of measuring the perturbation: the \( L_1 \) norm sums the absolute differences across coordinates, the \( L_2 \) norm measures the Euclidean distance, and the \( L_\infty \) norm considers the maximum absolute difference between any coordinate pair. These metrics enable precise control over the magnitude of perturbations, ensuring that the adversarial example remains effective while appearing indistinguishable from the original input.

In recent years, various adversarial attack techniques have been developed to exploit the vulnerabilities of ML models, including gradient-based methods \cite{goodfellow2014explaining, kurakin2018adversarial, madry2017towards, dong2018boosting, lin2019nesterov}, optimization-based methods \cite{DBLP:journals/corr/SzegedyZSBEGF13, carlini2017towards}, score-based methods \cite{ilyas2018black, li2019nattack} and decision-based methods \cite{brendel2017decision, chen2020boosting}. In this paper, we mainly focus on implementing the gradient-based adversarial attack methods: Fast Gradient Sign Method (FGSM) and Projected Gradient Descent (PGD). Gradient-based adversarial attacks are crafted by exploiting gradients of input data with respect to the target model's loss function, introducing perturbations that mislead model predictions. These attacks are widely used is classical ML for adversarial example generation.
\renewcommand{\thefigure}{S\arabic{figure}}  
\setcounter{figure}{0}

\begin{figure*}
\begin{center}
\includegraphics[width=1.0\linewidth]{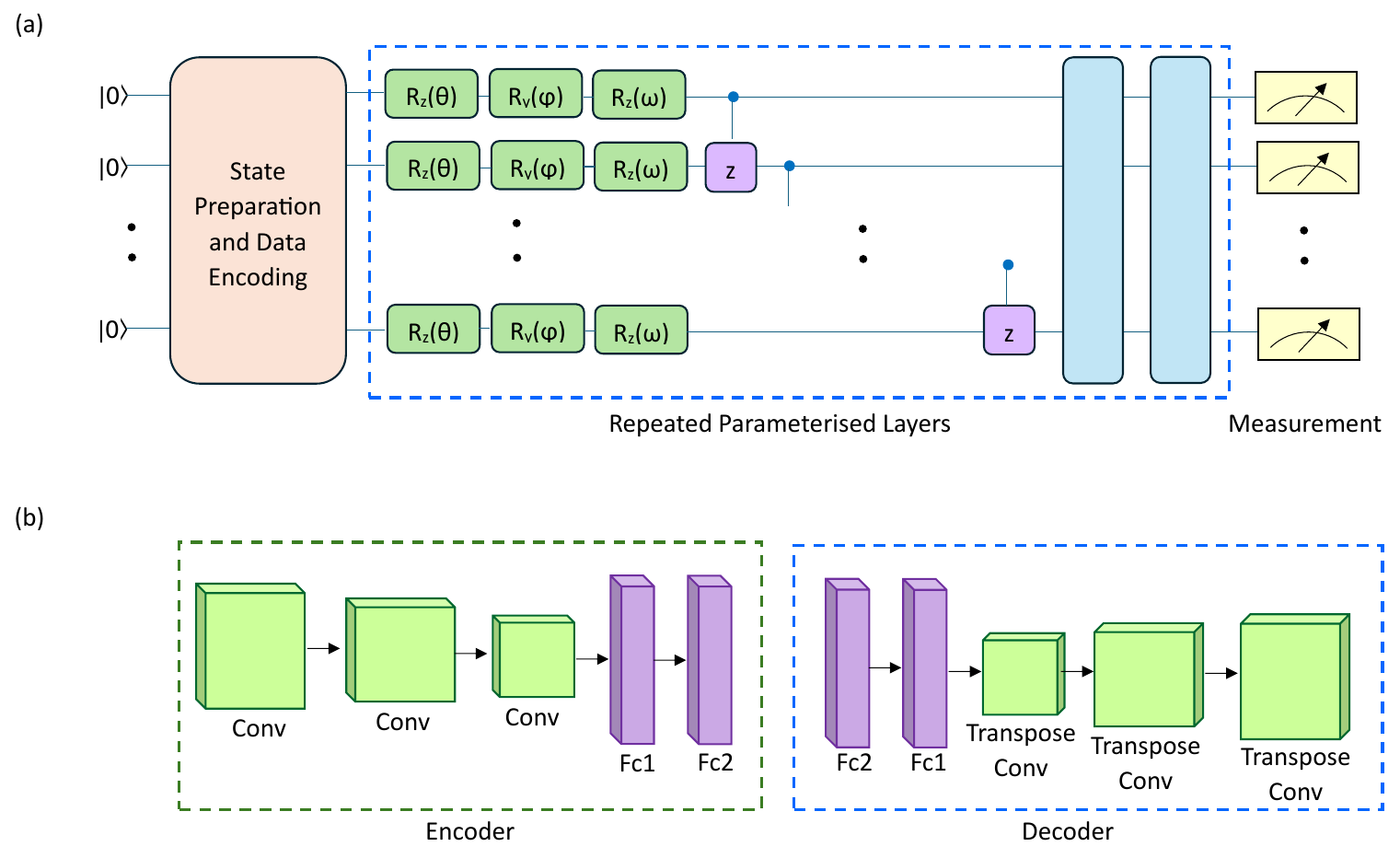}
\end{center}
   \caption{Architecture of the QVC and classical encoder-decoder circuit. In this Figure, (b) represents the QVC, which follows a three-step process: first, classical data is encoded into quantum states using amplitude encoding; next, the encoded data is processed through parameterised quantum layers consisting of single-qubit rotation gates and controlled-Z (CZ) gates; finally, a measurement layer extracts the classification result. The QVC is implemented with quantum circuit depths of 100 layers. (c) shows the architecture of the classical encoder-decoder network. The encoder compresses the input into a compact, low-dimensional latent representation using convolutional layers and fully connected (FC)layers. The decoder reconstructs the original input from this latent vector through fully connected and transposed convolutional layers, producing an output image of size $28\times28\times1$.}
\label{fig:overview}
\end{figure*}
FGSM \cite{goodfellow2014explaining} perturbs the input data in a way that maximises the model's prediction error. The idea behind FGSM is to compute the gradient of the loss for the input data and then perturb the data in the direction that increases the loss the most. For a classifier $f:X \rightarrow Y$, the FGSM attack generates adversarial sample $x^\wedge$ that is similar to the original training example $x$. It calculates the gradient of the loss function with respect to the input $x$ as,
\begin{align}
\Delta_x L(y,f(x;\theta)),
\label{eq:9}
\end{align}
where $y$ represents true label and $\theta$ represents model parameters. This gradient indicates how the loss changes with small variations in the input $x$. The adversarial example $x^\wedge$ is generated by perturbing the original input $x$ in the direction that maximizes the loss,
\begin{align}
x^\wedge = x + \epsilon \cdot \text{sign}(\nabla_x L(y, f(x; \theta))),
\label{eq:9}
\end{align}
Here, $\epsilon$ is a small positive constant that determines the magnitude of the perturbation. The 
$sign$ function extracts the positive or negative direction of the gradient, indicating the direction in which to perturb the input sample.

Unlike FGSM, which applies a single-step perturbation, PGD performs multiple gradient update steps to iteratively refine the adversarial examples. Each step uses the sign of the gradient of the loss with respect to the input to maximise the loss at each iteration. For input $x$ at step $t$, PGD attack can be formulated as:
\begin{equation}
   x^{t+1} = x^t + \alpha \cdot \text{sign}(\nabla_x L(f(x^t), y))
\end{equation}
where $\alpha$ is the step size, and $\nabla_x L$ denotes the gradient of the loss with respect to the input.
After each gradient step, PGD projects the perturbed input as:
\begin{equation}
   x^{t+1} = \text{clip}(x^{t+1}, x - \epsilon, x + \epsilon)
\end{equation}
This clipping function ensures that each updated  $x^{t+1}$ remains within an $\epsilon$-ball around the original input $x$, preventing excessive deviation from the original image.

\begin{figure*}
\begin{center}
\includegraphics[width=0.8\linewidth]{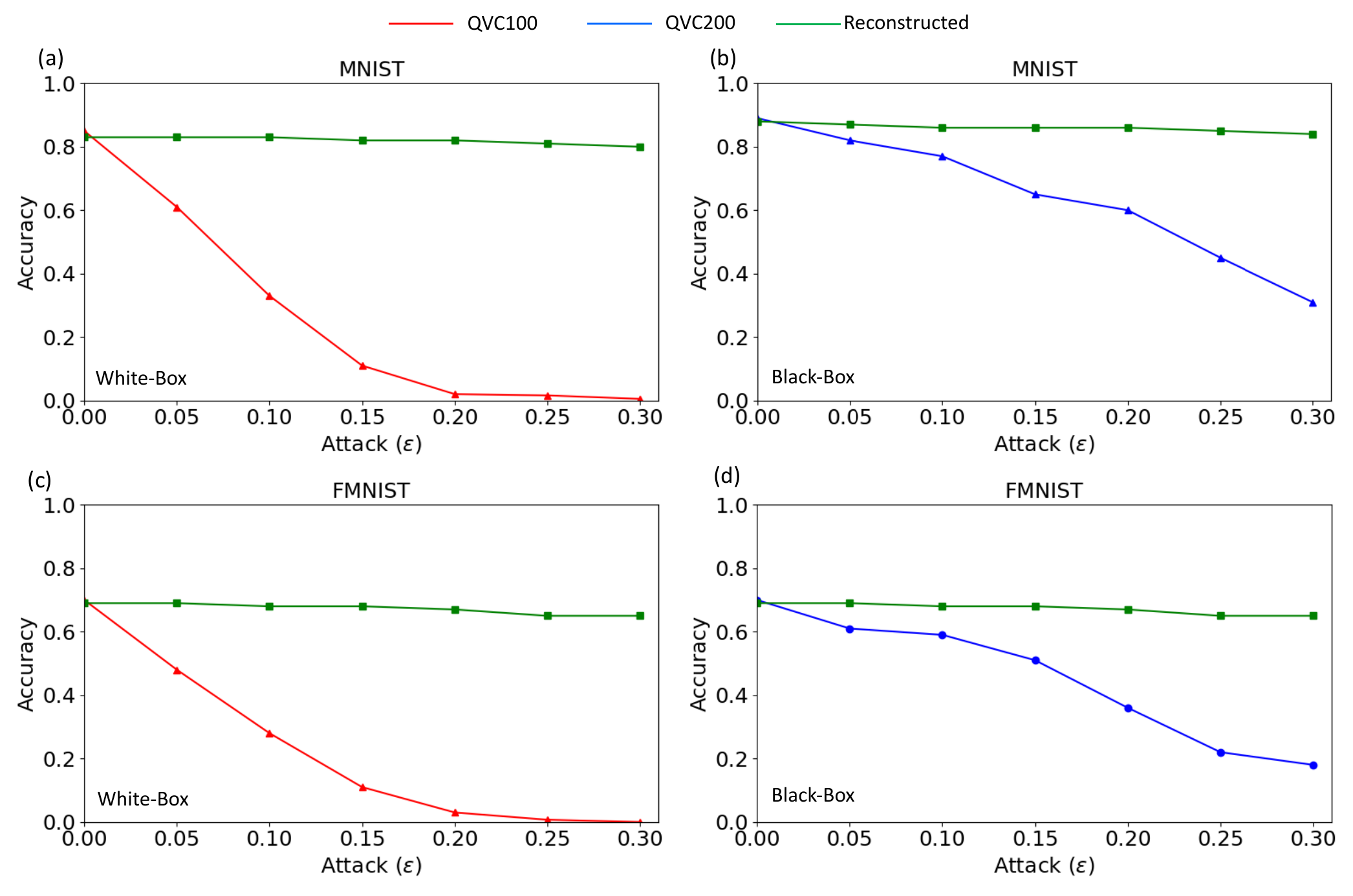}
\end{center}
   \caption{Plot of classification accuracy of quantum classifier under varying attack strengths, ranging from 0 to 0.3. For all the cases, the attacks are generated by QVC100 model. Panels (a), and (c) show the performance of quantum classifier under FGSM attacks for the MNIST and FMNIST datasets in white-box setting, respectively.  (b) and (d) represents the impact of FGSM attacks on the quantum classifier for the same datasets in black-box scenario. Across all cases, the classification accuracy decreases as the attack strength increases, highlighting the susceptibility of the models to quantum adversarial attacks. However, the results demonstrate a significant recovery in classification accuracy when the adversarial examples are passed through the classical autoencoder.}
\label{fig:result2}
\end{figure*}

\section{Adversarial Attacks on Quantum Variational Circuit}
\label{sup-QVC}
The adversarial examples are generated on a QVC using FGSM and PGD attacks. The architecture of QVC is shown in Fig. \ref{fig:overview} (b). Our QVC follows a three-step process: encoding classical data into quantum states, processing these quantum states through a variational quantum circuit and performing measurements for class label prediction. We use amplitude encoding \cite{larose2020robust}, which efficiently maps $28 \times 28$ grey-scale images from MNIST and Fashion-MNIST datasets into 10 qubits. In amplitude encoding, the classical data is mapped onto a quantum state, where each component of the classical data corresponds to the amplitude of the quantum state. For a classical dataset $x = (x_1, x_2,....,x_N)$, where $x_i$ is a normalized value, the quantum state can be written as:
\begin{equation}
    |\psi\rangle = \sum_{i=0}^{N} x_i |i\rangle,
\end{equation}
where $|i\rangle$ are the computational basis states and $N = {2^n}$ (where $n$ is the number of qubits). The encoded data is then passed through a trainable parameterised quantum circuit consisting of multiple variable layers. Each layer is composed of single-qubit rotation gates, followed by CZ gates between neighboring qubits to entangle the qubits within each layer. In our QVC, we employ 100 layers of quantum circuit. After passing through the quantum circuit, the Z-expectation value is measured for each qubit for class label prediction. The expected value for a given qubit can be represented as:
\begin{equation}
    \langle Z \rangle = \langle \psi(\theta) | Z | \psi(\theta) \rangle,
\end{equation}
where $Z$ is the $PauliZ$ operator, and $\theta$ denotes the parameters of the quantum circuit.

We compute the gradients of the QVC by differentiating the loss with respect to the input using a parameter-shift rule. The gradient of the loss function $L$ with respect to a parameter $\theta_i$ is given by:
\begin{equation}
\frac{\partial L}{\partial \theta_i} = \frac{1}{2} \big(L(\theta_i + \pi/2) - L(\theta_i - \pi/2)\big),
\end{equation}
These gradients quantify how changes in the input data affect the loss function to effectively craft  adversarial perturbations.

\section{Purifying Adversarial Examples using Classical Auto-Encoder}
\label{sup-enc}
Auto-encoders are extensively used for dimensionality reduction, latent feature learning, and generative models in classical ML. In this work, we use classical auto-encoder to reduce the adversarial perturbations from the attacked images. Here, the auto-encoder is working in a slightly different way compared to typical auto-encoders used for  representing data in lower-dimensional space (latent feature space). The auto-encoders learn to capture only useful information from input data (ignoring the perturbations) by changing the reconstruction criterion. QVC generated adversarially perturbed images are passed to the encoder as input. The encoder extract meaningful features using multiple convolutional operations, and these latent features are then passed to a decoder. The decoder consists of a set of deconvolutional layers to reconstruct the original images. The detail architecture of the auto-encoder is illustrated in Fig. \ref{fig:overview} (b). The encoder takes perturbed images of size $28 \times 28\times1$ as input, then the first convolutional layer applies $16$ filters with a $3 \times 3$ kernel, a stride of 2, and  padding of 1. The second  convolutional layer consists of $32$ filters with the same kernel size, stride, and padding as the first convolutional layer, resulting in an output of size $32 \times 7 \times 7$. This output is then flattened into a $1D$ tensor and passed through a fully connected layer, reducing it to a compact latent representation of size 20. The decoder reconstructs the image from this 20-dimensional latent vector. It begins with a fully connected layer that expands the latent vector back to a size of  $32 \times 7 \times 7$, followed by reshaping into a 3D tensor. The upsampling is performed using two transposed convolutional layers and generates the final reconstructed image of size $28 \times 28\times1$. ReLU activation is applied in all intermediate layers, while a Sigmoid activation in the final layer ensures the output pixel values are normalised between 0 and 1.

Mathematically, the encoder maps the adversarial image $x_adv$ into a low-dimensional latent representation $z \in \mathbb{R}^{20}$. Initially, the perturbed images are passed through convolutional layers:
\begin{equation}
x_i = \text{Conv2D}(x_{i-1}, W_i, b_i), \quad i = 1, 2, \ldots, n,
\end{equation}
where \( n \) is the number of convolutional layers, and \( W_i, b_i \) are the weights and biases of the \( i \)-th layer. The output of the last convolutional layer is a feature map $x_2 \in \mathbb{R}^{32 \times 7 \times 7}$, and then flattened into a vector $x_f \in \mathbb{R}^{1568}$, followed by a fully connected layer:
\begin{equation}
z = \sigma(W_{\text{fc}} x_f + b_{\text{fc}}),
\end{equation}
where \( W_{\text{fc}}, b_{\text{fc}} \) are the weights and biases of the fully connected layer, and $\sigma$ is the activation function.

The decoder maps the latent vector $z$ to a higher-dimensional feature map:
\begin{equation}
  x_d = \phi(W_{\text{fc}}' z + b_{\text{fc}}'),  
\end{equation}
where \( W_{\text{fc}}', b_{\text{fc}}' \) are the weights and biases of the fully connected layer in the decoder. The reshaped feature map $x_d$ is upsampled using a transposed convolution layer:
\begin{equation}
x_j = \text{ConvTranspose2D}(x_{j-1}, W_j, b_j), \quad j = 1, 2, \ldots, m,
\end{equation}
where \( m \) is the number of transposed convolutional layers, and \( W_j, b_j \) are the weights and biases of the \( j \)-th layer. The output of the last transposed convolutional layer is normalised by Sigmoid activation function to ensure that the values of the reconstructed images $\hat{x}$ are in [0, 1].
